\documentclass[lettersize,journal]{IEEEtran}

\usepackage{pbox}
\usepackage{generic}
\usepackage{cite}
\usepackage{amsmath,amssymb,amsfonts}
\usepackage{algorithmic}
\usepackage{graphicx}
\usepackage{algorithm,algorithmic}
\usepackage{hyperref}
\hypersetup{hidelinks=true}
\usepackage{textcomp}
\usepackage{graphicx}
\usepackage{subcaption}

\begin{document}

\title{A Nonlinear Analysis Software Toolkit for Biomechanical Data}

\author{
\IEEEauthorblockN{Shifat Sarwar, Aaron Likens, Nick Stergiou,}
\and
\IEEEauthorblockN{Spyridon Mastorakis}\thanks{S. Sarwar is with the Department of Computer Science, University of Nebraska at Omaha, Omaha, NE, USA (email: ssarwar@unomaha.edu). A. Likens and N. Stergiou are with the Department of Biomechanics, University of Nebraska at Omaha, Omaha, NE, USA (emails: alikens@unomaha.edu, nstergiou@unomaha.edu). S. Mastorakis is with the Department of Computer Science and Engineering, University of Notre Dame, Notre Dame, IN, USA (email: mastorakis@nd.edu).}
}



\setlength{\skip\footins}{4pt}

\maketitle

\begin{abstract}
In this paper, we present a nonlinear analysis software toolkit, which can help in biomechanical gait data analysis by implementing various nonlinear statistical analysis algorithms. The toolkit is proposed to tackle the need for an easy-to-use and friendly analyzer for gait data where algorithms seem complex to implement in software and execute. With the availability of our toolkit, people without programming knowledge can run the analysis to receive human gait data analysis results. Our toolkit includes the implementation of several nonlinear analysis algorithms, while it is also possible for users with programming experience to expand its scope by implementing and adding more algorithms to the toolkit. Currently, the toolkit supports MatLab bindings while being developed in Python. The toolkit can seamlessly run as a background process to analyze hundreds of different gait data and produce analysis outcomes and figures that illustrate these results. 
\end{abstract}

\begin{IEEEkeywords}
Biomechanics, human gait, nonlinear gait data analysis, software toolkit
\end{IEEEkeywords}

\section {Introduction}
\label{sec:intro}

Biomechanics is a field that studies movement patterns of the human and animal bodies~\cite{knudson2007fundamentals}. Gait data, which is essentially the collection of data from an individual's movement patterns during walking, is a crucial component for understanding human movement to answer clinical questions. Gait data can be analyzed to provide valuable insights on various aspects of movement, such as diagnosis and monitoring of conditions related to human movement, assessment of the effectiveness of movement exercises, development of assistive devices, and identification of unique individual patterns~\cite{stergiou2004innovative, likens2021irregular}. Nonlinear statistical analysis techniques have proven to be effective in studying human gait since linear methods are often inadequate and/or inapproriate for evaluating the temporal organization and the complexity of human movement~\cite{stergiou2018nonlinear}. 

While there are several nonlinear analysis techniques available for the analysis of gait patterns, the process of selecting and utilizing these techniques can be time-consuming and laborious. Researchers often need to collect, modify, and implement multiple algorithms for different analyses, resulting in the duplication of work across multiple research projects. 
To address this challenge, in this paper, we present the development of a software toolkit that contains several nonlinear analysis techniques and a user-friendly interface, enabling researchers with minimal technical and software expertise to analyze gait data effectively. Currently, our toolkit supports MatLab bindings while being developed in Python. Furthermore, it is also possible for users with programming experience to expand its scope by implementing and adding more algorithms to the toolkit. 

Our contribution in this paper is two-fold: (i) we present the design and development process of our toolkit along with the nonlinear analysis techniques that we have implemented; and (ii) we evaluate the performance of our toolkit in terms of execution time and memory consumption for all implemented algorithms. The rest of the paper is structured as follows. In Section~\ref{sec:back}, we discuss prior related work. In Section~\ref{sec:design}, we present the design of our software toolkit. In Section~\ref{sec:eval}, we present our toolkit's evaluation. Finally, in Section~\ref{sec:conclusion}, we conclude our paper. 



\section{Related Work}
\label{sec:back}



\subsection {Biomechanics Data Analysis Toolkits} 

Gait data is captured and utilized in many formats but mostly in C3D formats. Barre \textit{et al} built a toolkit that assisted in biomechanical data visualization and processing. The authors in that paper built an open-source framework that reads C3D files (which are normally associated with biomechanical movement data) and allows the users to easily read, modify and process the data. 
This toolkit mostly handles different data formats and allows data conversion and filtering. It allows users to calculate variables and conduct kinematic and kinetic analyses, electromyography (EMG) data processing and visualize the data in 2D and 3D plots~\cite{barre2014biomechanical}. Some of its features are also seen in another tool called the OpenSim by Seth \textit{et al}, which allows musculoskeletal modelling and simulation~\cite{seth2011opensim}. Both tools utilize MatLab and Python for data analysis and visualization.

The Vicon toolkit for motion capture and gait analysis was introduced by Goldfarb \textit{et al}~\cite{goldfarb2021open}. 
The analysis conducted via this toolkit is different from ours. The main analysis the Vicon toolkit allows users to do is to extract marker trajectories, create and manipulate rigid bodies, separate gait cycles, examine force plate data, filter EMG data, and learn motion models. With Vicon the data can be preprocessed with Python and MatLab packages. Another open-source toolkit has been proposed, which works as a python program package for conducting generic biomechanical analysis of human motion data~\cite{chenier2021kinetics}.

All the toolkits mentioned have some commonalities, including the availability and usage of Python and MatLab bindings, as well as the fact that they allow their users to integrate custom codes for their analysis easily. These features are also supported in our toolkit.

\subsection {Gait Data Analysis Techniques}

Chau \textit {et al.} explored different analysis techniques for gait data using fuzzy, multivariate statistical and fractal methods~\cite{chau2001review}. Their paper is divided into two parts: the first explores the techniques using fuzzy systems that include clustering and coding, which help address the variability of gait data by bringing randomness based on probability and help show patterns in the structure that could have been disregarded. The multivariate statistical approaches in the study include Principal Component Analysis (PCA), Factor Analysis (FA), and Multiple Correspondence Analysis (MCA). PCA can mainly help give insight into and interrelationship of different walking strategies, identify important gait patterns from the limb movement data, study the entire temporal gait waveform, and find out if there is pathological deviation following the complete gait. FA can give insight into muscle activation patterns from the data. Performing MCA on the gait data can show numerous gait variables at the same time by representing both the linear and nonlinear relationships as graphs. Lastly, Detrended Fluctuation Analysis (DFA) and Relative Dispersion (RD) analysis are examples of fractal analysis. These techniques measure the long-range dynamics of gait data. Their usage can discover patterns from temporal fluctuations in human gait data. They are also useful to perform quantitative assessments.
The second part of the paper by Chau \textit {et al.} explored the effects of neural networks and wavelet transformation on gait data~\cite{chau2001review2}. The use of neural network models on gait data can help analyze large amounts of data, perform research on relationships that are difficult to manage otherwise, manipulate the high dimensionality in gait data, map the data nonlinearly and capture temporal dependency.

However, these techniques are all difficult to perform and study without the help of proper software capable of performing all the calculations for the analysis and displaying the results for the user to accelerate research in this field. One of the analysis techniques, DFA for fractal analysis, discussed above, is already available for gait data analysis with our toolkit, among many others. More experienced users can also add other analysis techniques and expand the scope of the toolkit for biomechanical research. 

\subsection {Nonlinear Analysis}

Nonlinear analysis techniques are often more complex to understand and use than linear analysis. The lack of understanding could also be due to lack of knowledge of the benefits of these techniques or the true limitations of linear methods which, however, are taught in every introductory statistical class. Furthermore, without proper toolkits, these techniques can be difficult to implement and use repeatedly. To this end, such toolkits have been in development for a long time. Parker \textit{et al} introduced INSITE, a software-based solution for analyzing nonlinear dynamical systems~\cite{parker1987insite}. Their toolkit contains algorithms that support the calculation and representation of trajectories, bifurcation diagrams, and phase portraits, along with calculating the Lyapunov exponents of an autonomous system. Similar tools in recent years include a toolkit for nonlinear heart rate variability analysis by Garcia \textit{et al}~\cite{garcia2013software}. The authors present an open-source package for R language platforms. With the help of their nonlinear analysis toolkit, they could generate an automated estimation of embedding dimension and time lag for phase space reconstruction, noise reduction, and algorithms for many nonlinear analysis techniques needed to analyze heart rate data. Their tool can also check the nonlinearity of the provided heart rate data before conducting any analysis to determine the validity of their result using a method called the Keenan test and neural networks. The most significant analysis algorithms included in the package give results for the maximum Lyapunov exponent, generalized correlation dimension, sample entropy, Recurrence Quantification Analysis (RQA), DFA, and  Poincaré Plots. This toolkit focuses on processing heart rate data. On the other hand, our toolkit is designed to prepare (preprocess) gait data for nonlinear analysis and analyze captured data gait. The algorithms for getting the Lyapunov exponent, RQA, and DFA are also available in our toolkit to conduct nonlinear analysis on gait data. Feeding gait data on the heart rate variability toolkit and heart rate data on the gait data analysis toolkit is not feasible due to the variations in the collected data, rendering these algorithms unusable. However, it may be possible to achieve this by modifying the data processing techniques utilized in both toolkits.
\section{Toolkit Design and Development}
\label{sec:design}



\subsection{Overview}

We selected Python 3 as the programming language for the development of the toolkit. This decision was based on the language's clean syntax and ease of code modification. The availability of powerful libraries such as pandas, numpy, and Scipy made Python 3 our preferred choice for gait analysis. 
To analyze gait data, the initial step in any framework is to appropriately load and pre-process the input data before actual processing. The dataset used by our toolkit consists of input files containing gait data spread across numerous columns, each representing data collected from a specific attribute of the human body in various axes and degrees. Accurate attribute naming in the dataset is critical, as most algorithms in the framework operate on the data of a single column representing a certain part of the body of a human subject. The framework can automatically detect the file type and read the data from the files accordingly, supporting CSV and parquet file formats. The framework is designed to run algorithms continuously and generate results for each attribute column listed. The output results are generated in separate text files, each representing a distinct algorithm output. The framework's implementation in Python 3 utilizes the pandas' package to transform input column data into a DataFrame, which serves as input for the various available algorithms.

The toolkit provides users with the flexibility to customize their analysis by allowing them to set specific parameters, choose from available algorithms or use default settings. It has been developed with open-source principles in mind, providing proper documentation about its usage and code, and is readily accessible for anyone to use from GitHub~\cite{sarwar2023}. Its open-source nature encourages community involvement and allows for further development, testing, and optimization.

\begin{figure}
\centering
\includegraphics[width=1\linewidth]{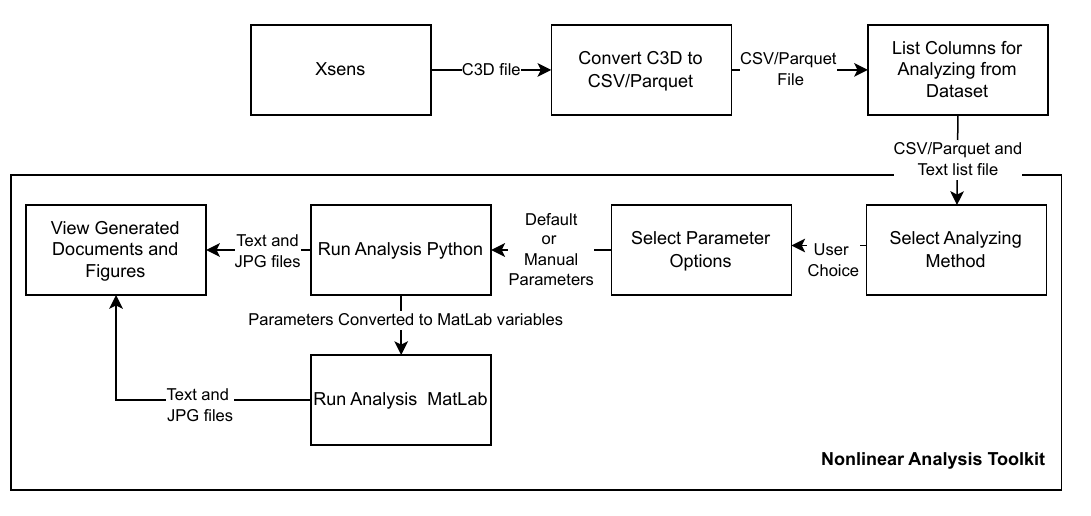}
\captionsetup{justification=justified}
\caption{Framework Data Flow Diagram}
\label{Figure:dataflow}
\vspace{-0.6cm}
\end{figure}

Figure~\ref{Figure:dataflow} illustrates the data flow process, which commences with the collection of kinematic and kinetic data using Inertial Measurement Unit (IMU) sensors strategically placed on various parts of the human body. This data is then gathered by Noraxon software, which generates an output that can be exported into C3D file formats. Subsequently, the exported file can be converted to CSV or Parquet files, which are compatible with our software toolkit. To streamline the analysis process, the user needs to specify the attributes they wish to be analyzed from the file. Given the potential for many attributes to be processed, it is recommended that the user provides a list of these attributes in a separate text file. By adopting this approach, the user can effectively manage the data and ensure that the analysis is performed on the relevant attributes.

\subsection{Implemented Algorithms}

The nonlinear analysis toolkit offers a comprehensive set of algorithms for analyzing gait data, providing researchers with valuable insights into various attributes of the data. The toolkit includes several popular nonlinear analysis methods, including the Average Mutual Information (AMI) algorithm that can generate the time lag from the time series data for state space reconstruction, and the False Nearest Neighbor (FNN) algorithm that provides the embedding dimension results. It also calculates results for Refined Composite Multiscale Sample Entropy, Multiscale Entropy, Multiscale Fuzzy Entropy, and Generalized Multiscale Entropy utilizing the work by  Wu et al. 2014\cite{wu2014analysis}, Permutation Entropy from Bandt and Pompe, 2002 \cite{bandt2002permutation}, Sample Entropy from Richman and Moorman, 2000\cite{richman2000physiological}, and Symbolic Entropy from Aziz and Arif, 2006\cite{aziz2006complexity}. It can also compute the Cross Approximate Entropy between two time series using the method from Richman and Moorman, 2000\cite{richman2000physiological}. Popular nonlinear analysis methods like Detrended Fluctuation Analysis (DFA) developed from studies by Damouras et al. 2010\cite{damouras2010empirical}, Mirzayof and Ashkenazy, 2010\cite{mirzayof2010preservation}, and Peng et al. 1995\cite{peng1995quantification} and Recurrence Quantification Analysis (RQA) using methods presented in the study by Wallot et al. 2016\cite{wallot2016multidimensional} and Eroglu et al. 2014.\cite{eroglu2014entropy} are also included. Furthermore, the toolkit can be used to calculate the Lyapunov Exponent values using two different methods following Rosenstein et al. 1992\cite{rosenstein1993practical} and Wolf's 1985 paper\cite{wolf1985determining} respectively. 
The toolkit also automates the preprocessing steps required for data analysis, saving researchers time and effort. Furthermore, the algorithms can be customized by adjusting the parameter settings to obtain different results, making the toolkit versatile and adaptable to different research needs. 

\begin{figure}
\centering
\captionsetup{justification=justified}
\includegraphics[width=1\linewidth]{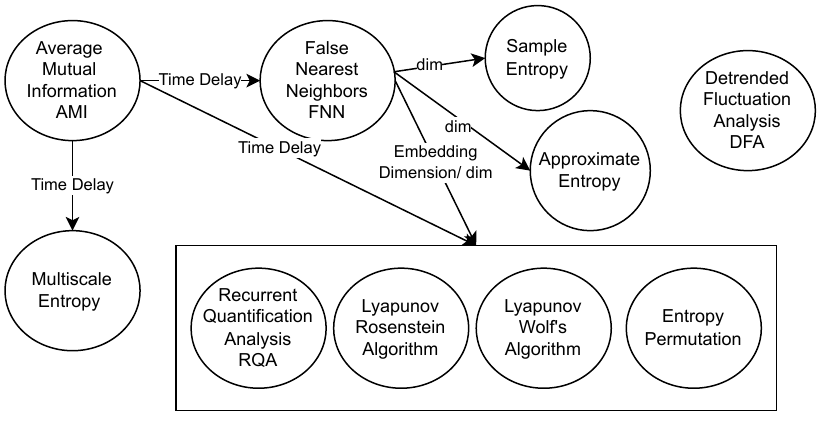}
\caption{Algorithms and their relations}
\label{Figure:algo_relations}
\vspace{-0.6cm}
\end{figure}

The nonlinear analysis toolkit presented here offers a user-friendly interface and automation of the preprocessing steps required for gait data analysis. One notable feature of the toolkit is its ability to automatically generate input parameters like time delay and embedding dimension for the analysis algorithms, eliminating the need for manual inputs from the user. A visualization of the relationships among the different implemented analysis algorithms can be seen in Figure~\ref{Figure:algo_relations}. 

\subsection{Language Bindings}

The nonlinear analysis toolkit is compatible with both Python 3 and MatLab, providing users with the flexibility to run code written in either language. MatLab versions of the code typically require fewer code conversion steps than Python, making them slightly faster for data analysis. Additionally, MatLab is well-suited for heavy mathematical calculations, making it an ideal choice for certain algorithms. To support MatLab codes, users must have the MatLab program installed on their system. Most algorithms in the toolkit are available in both MatLab and Python versions. Adding new analysis algorithms in either language is straightforward and would only require the addition of a single function call statement.

Certain analysis methods, such as RQA and DFA, utilize MatLab libraries to generate their outputs. In such cases, users must use the MatLab version of the toolkit. However, if they prefer, they may also choose to use the Python 3 version, though this will require more time and memory for running the analysis that may not be available. While all algorithms could have been implemented using MatLab, we chose to include a Python 3 version to increase accessibility for users. Python 3 does not require users to purchase any additional software packages and is simpler to install and use. We aim to provide users with a seamless and customizable toolkit that can accommodate their specific needs. By offering compatibility with both Python and MatLab, we can make the toolkit more accessible to a wider range of users.

\subsection{Error Handling}

Our software toolkit processes input data generated from human gait, which can be easily converted into a ‘CSV’ or ‘Parquet’ dataset from the original C3D file format. To ensure seamless processing, users are required to enter only minimal information. The algorithms embedded within the toolkit have been carefully designed to accommodate most datasets, except in cases where there are issues with a particular column of time series data or where the wrong algorithm has been selected. However, in such situations, the user will be promptly notified, and they can correct their choice of dataset or algorithm accordingly.

The toolkit is compatible with most computer systems supporting MatLab and Python code implementations, except for the RQA algorithm, which requires significantly larger memory compared to other algorithms available in the toolkit. From table~\ref{Table:datasets}, it is evident that analyzing using RQA will result in the most time and resources. Instances where the user gives the radius and is not calculated before conducting RQA lower the average time to about 2400s per execution, but the space usage remains the same. The generated result array from RQA can also take up gigabytes of space to store the generated matrix representing the generated graph.

\subsection {Optimizations and Multi-Threading}

The purpose of our toolkit is to analyze large volumes of data using multiple algorithms directed on specific columns representing different axes of the movement gathered from different body parts during human walking. Speed and accuracy are key considerations in this process, and we have taken great care to ensure that the algorithms generate results rapidly and in a clear and organized manner that does not create confusion for users when they read through the generated documents. To ensure accuracy, our toolkit incorporates algorithms that have been extensively tested and validated by the Department of Biomechanics at the University of Nebraska at Omaha and have been used for the analysis of human gait data collected in the context of studies over the years. To ensure speed, we limited the number of preprocessing steps before feeding the data to an algorithm. Specifically, we have implemented a streamlined preprocessing approach that minimizes the number of steps required before feeding the data into the algorithms. As an example, the movement of a single person over 200 seconds can generate up to 50,000 rows of data, which requires sufficient space for a computer to execute the algorithms effectively. Considering the analysis of over 250 columns of data to obtain the DFA results for a human subject's movement in all axes and degrees of moving, such analysis would typically take approximately 10,000 seconds or roughly 3 hours.

To address this issue, our toolkit allows users to launch multiple instances of the program (threads), thereby enabling multi-threading to accelerate the data analysis process~\cite{mastorakis2021networking}. The number of executions required depends on the capacity of the user's system, and we recommend that users take advantage of this feature to optimize their analysis time.

\subsection{Documentation}

Our software toolkit comes with a detailed manual that has been thoughtfully curated to cater to the needs of both computer scientists and biomechanists. The manual provides comprehensive guidance on executing the various operations as required, making it a valuable resource for users with varying levels of technical expertise. In addition, we have provided detailed instructions on how to modify, update or add algorithms to the toolkit to ensure that future users can benefit from the development of this open-source product.

To add a new statistical analysis algorithm, users simply need to push the code into the respective file containing Python or MatLab software, following the instructions provided. Specifically, the user needs to add the calling function in the ``driver.py'' and ``functions.py'' scripts. For added convenience, we have included an ``other algorithm'' option, allowing users to select from the implemented codes or add the name of the algorithm by making changes only to the ``driver.py'' script.

\section {Evaluation}
\label{sec:eval}


\begin{table*}[ht]
\caption{Algorithm Results and Resource Utilization}
\vspace{-0.2cm}
\label{Table:datasets}
\begin{tabular}{|c|c|c|c|}
\hline
\textbf{Algorithm}                       & \textbf{Generated Results}                                                                                                                                                                                                              & \textbf{\begin{tabular}[c]{@{}c@{}}Average Time for \\ Single Execution (s)\end{tabular}} & \textbf{\begin{tabular}[c]{@{}c@{}}Average Memory for \\ Single Execution (GB)\end{tabular}} \\ \hline
Detrended Fluctuation Analysis (DFA)     & Detrended Fluctuation Analysis Scaling Exponent                                                                                                                                                                                         & 39.84                                                                                     & 0.07                                                                                         \\ \hline
Average Mutual Information (AMI)         & Average Mutual Information between two Arrays                                                                                                                                                                                           & 8.39                                                                                      & 0.26                                                                                         \\ \hline
False Nearest Neighbor (FNN)             & Embedding Dimension                                                                                                                                                                                                                     & 391.40                                                                                    & 0.26                                                                                         \\ \hline
Sample Entropy (Ent\_Samp)               & Sample Entropy                                                                                                                                                                                                                          & 142.48                                                                                    & 0.21                                                                                         \\ \hline
Approximate Entropy (Ent\_Ap)            & Approximate Entropy                                                                                                                                                                                                                     & 512.28                                                                                    & 0.27                                                                                         \\ \hline
Permutation Entropy (Ent\_Permu)         & Permutation Entropy                                                                                                                                                                                                                     & 47.19                                                                                     & 0.27                                                                                         \\ \hline
Symbolic Entropy (Ent\_Symbolic)         & Normalized Corrected Shannon Entropy                                                                                                                                                                                                    & 0.23                                                                                      & 0.27                                                                                         \\ \hline
Multi-scale Entropy (Ent\_MS\_Plus)      & \begin{tabular}[c]{@{}c@{}}Refined Composite Multiscale Entropy (RCMSE), \\ Composite Multiscale Entropy (CMSE), \\ Multiscale Entropy (MSE), \\ Multiscale Fuzzy Entropy (MSFE), \\ Generalized Multiscale Entropy (GMSE)\end{tabular} & 673.22                                                                                    & 0.20                                                                                         \\ \hline
Recurrence Quantification Analysis (RQA) & Recurrence Plot (RP) matrix                                                                                                                                                                                                             & 3979.47                                                                                   & 7.94                                                                                         \\ \hline
Lyapunov Rosenstein (LyE\_R)             & Short/Local/Long/Orbital Lyapunov Exponent                                                                                                                                                                                              & 311.63                                                                                    & 0.27                                                                                         \\ \hline
Lyapunov Wolf's Algorithm (LyE\_W)       & Largest Lyapunov Exponent                                                                                                                                                                                                               & 11.08                                                                                     & 0.27                                                                                         \\ \hline
\end{tabular}
\end{table*}


In this section, we present the evaluation results that we have collected for our toolkit while evaluating each individual algorithm in terms of execution time and memory usage. We evaluated our toolkit on a server equipped with an Intel Xeon Silver 4216 CPU @2.10GHz and 64 GB of memory. In our evaluation process, we utilized gait datasets collected from studies conducted by the Department of Biomechanics at the University of Nebraska at Omaha. 

We present our results in Table~\ref{Table:datasets}. Our results demonstrate that less complex implemented algorithms can be executed in general faster than more complex ones. Moreover, less complex algorithms also consume smaller amounts of memory. Based on our results, the RQA algorithm is the most computationally heavy, thus it takes the longest to execute and consumes the largest amount of memory. The RQA algorithm loads the gait data and iterates over it many times to generate the recurrence graph. Furthermore, it takes about 1000 seconds extra to determine the radius value from a given recurrence value. The 1000 seconds can be cut short if the user provides the radius value instead of the recurrence as the parameter. 

Other than RQA, all the algorithm functions consume roughly the same amount of memory (about 0.20-0.27 GB).  
Each algorithm results in a different execution time though. The Entropy Multi-Scale (MS) Plus  algorithm results in the second-highest execution time of about 673 seconds on average, which can change depending on the parameters supplied. The approximate entropy algorithm results in the third-highest execution time of above 500 seconds on average. The FNN algorithm consumes about 391 seconds per execution and is followed by the Lyapunov Exponent algorithm, using the Rosenstein method~\cite{rosenstein1993practical}, which results in 311 seconds per execution on average. Sample entropy values can be generated within 142 seconds, while all other algorithms result in an average of 8 to 50 seconds per execution.

\section {Conclusion and Future Work}
\label{sec:conclusion}

This article presents a nonlinear statistical analysis toolkit designed for analyzing human gait data. The toolkit offers a simple interface that can be easily operated by individuals with diverse levels of programming knowledge. It enables the analysis of hundreds of gait patterns using various statistical techniques without requiring users to have an in-depth understanding of coding or modifying complex nonlinear analysis algorithms. The toolkit also allows users to select default algorithm parameters or manually enter their parameters, offering flexibility in analysis. As an open-source product, it encourages the community to contribute to the development of the tool, allowing for continuous updates and improvements in functionality. 

Moving forward, there is a need to expand the toolkit by adding more nonlinear analysis techniques and developing a direct conversion mechanism to allow for the use of data directly from C3D files without the need for converter software. The RQA algorithm has some limitations in terms of storage and time usage, and further optimization will be investigated to reduce resource utilization when handling large amounts of data rows. A more advanced graphical user interface can also be added to provide an easier and more user-friendly selection of parameters and techniques. With these improvements in mind, our toolkit can continue to evolve and remain a valuable tool for analyzing gait patterns and collecting results in the field of biomechanics.

\section*{Acknowledgments}

This work is partially supported by the National Science Foundation (awards CNS-2104700, CNS-2306685, CNS-2016714, and CBET-2124918), ACM SIGMOBILE, and the National Institutes of Health (award NIGMS/P20GM109090).

\bibliographystyle{unsrt}
\bibliography{refs}

\end{document}